\documentstyle[12pt]{article}

\date{July 1995}
\author{
\bf Juan Carlos D'Olivo \\
Instituto de Ciencias Nucleares\\
Universidad Nacional Aut\'{o}noma de M\'{e}xico\\
Apartado Postal 70-543, 04510 M\'{e}xico, D.F., M\'{e}xico\\
\and
\bf Jos\'e F. Nieves\\
Laboratory of Theoretical Physics\\
Department of Physics, P. O. Box 23343\\
University of Puerto Rico\\
R\'{\i}o Piedras, Puerto Rico 00931-3343\\
\and
\bf Palash B. Pal \\
Indian Institute of Astrophysics,
Bangalore 560034, INDIA
}
\title{\bf Cherenkov radiation by massless neutrinos}

\begin{document}

\maketitle

\begin{abstract}
Due to their weak interactions, neutrinos can polarize a medium
and acquire an induced charge. We consider the Cherenkov
radiation emitted by neutrinos due to their effective electromagnetic
interactions  as they pass through a polarizable medium.
The effect exists even for massless, chiral neutrinos, where no physics
beyond the standard model needs to be assumed.
\end{abstract}

The study of the electromagnetic properties of neutrinos,
in vacuum as well as in a medium, has been a subject
of great interest over the years because of its
intrinsic interest and also because of its potentially
important consequences in a variety of physical, astrophysical
and cosmological contexts. Recently, several authors have
considered the Cherenkov radiation emitted by neutrinos
as they pass through a medium \cite{grimus}, and also the transition
radiation produced when they cross the interface between
two media with different dielectric properties \cite{sakuda,GrNe95}.
In these works, 
the authors have assumed that the neutrino has an intrinsic
magnetic and/or electric dipole moment (and hence also a mass),
which are responsible for the electromagnetic interactions
with the medium.

In a paper by two of the present authors \cite{indcharge},
it was pointed out that neutrinos acquire an induced
charge as they propagate through a medium as a consequence
of their weak interactions with the background particles \cite{oraevsky}.
That observation was based on the 1-loop
calculation of the effective electromagnetic vertex
of the neutrino, which was performed using the
methods of ``Quantum Statistical Field Theory''\footnote{
We prefer this name to the more often used ``Finite Temperature
Field Theory'', since the methods apply also
for zero temperature but finite density.} \cite{dnp}. The effective
charge was found to be nonvanishing even for a massless neutrino, where
no physics outside the standard model needs to be assumed.

Here we point out that, because of the induced electromagnetic
interactions of the neutrinos, they can emit Cherenkov
radiation (and transition radiation in the case that they
cross the interface between two media), even if they are massless
and do not have intrinsic electromagnetic dipole moments.
In this article we consider these effects, following a method
similar to that of Ref.~\cite{indcharge}, and using
the results of Ref.~\cite{dnp} as the basis of the calculations.

\section{Kinematics}
Our aim is to calculate the rate for the process
\begin{equation}\label{process}
\nu(k)\rightarrow \nu(k^\prime) + \gamma(q)\,,
\end{equation}
where, in the frame where the medium is at rest,
\begin{equation}\label{numomentum}
k^\lambda  =  (\omega,\vec K), \qquad
k^{\prime\lambda} = (\omega^\prime,\vec K') \,,
\end{equation}
denote the initial and final momenta of the neutrino, and
\begin{equation}\label{phmomentum}
q \equiv k - k' = (\Omega,\vec Q)
\end{equation}
denotes the momentum of the emitted photon.  Since we are interested
in the contribution to the Cherenkov radiation due to the
effective $\nu\nu\gamma$ interaction, for our purposes it is
sufficient to consider the case in which the neutrinos are strictly
massless in the vacuum.  Thus, we assume that the on-shell conditions
for the neutrinos is~\cite{footnote1}
\begin{equation}\label{nuonshell}
k^2 = 0, \qquad k'^2 = 0\,.
\end{equation}
Using $k'=k-q$, the second equation can be rewritten in the form
\begin{eqnarray}
\cos \theta \equiv {\vec K \cdot \vec Q \over KQ} = {2
\omega\Omega - \Omega^2 + Q^2 \over 2KQ} \,.
\label{cos}
\end{eqnarray}
Since $-1\leq \cos\theta \leq 1$, this implies
\begin{eqnarray}
\omega\Omega - KQ \leq {1\over 2} (\Omega^2 - Q^2) \leq
\omega\Omega + KQ \,.
\label{coslimits}
\end{eqnarray}
It is easy to see that these conditions cannot be satisfied for
$Q<\Omega$. Thus, we must have
\begin{eqnarray}\label{cerenkcond1}
Q > \Omega
\end{eqnarray}
in which case the right inequality of Eq.\ (\ref{coslimits})
is automatically satisfied, while the left inequality implies the
subsidiary condition
\begin{eqnarray}\label{cerenkcond2}
K Q - \omega \Omega >  0 \,.
\end{eqnarray}
Since we are assuming the vacuum on-shell relation for the
neutrino ($\omega = K$), the second condition is equivalent to the
first one.

The condition in Eq.\ (\ref{cerenkcond1}) shows that
for the photon we cannot take the vacuum disperstion relation.
Rather, it is important that we take into
account that its energy-momentum relation in an isotropic
medium is given by the solutions of
\begin{equation}\label{phonshelltr}
\frac{Q^2}{\Omega^2} = \varepsilon_t(\Omega,Q) \pm \varepsilon_p(\Omega,Q)
\end{equation}
for the transverse modes, and
\begin{equation}\label{phonshelllong}
\varepsilon_l(\Omega,Q) = 0
\end{equation}
for the longitudinal one.  The functions $\varepsilon_{t,l,p}$ are
the components of the dielectric response function of the system,
which are related to the components of the photon self-energy
by~\cite{pandcpodd}
\begin{equation}\label{eq6.24}
1 - \varepsilon_t = \pi_T/\Omega^2\,, \qquad 1 - \varepsilon_l = \pi_L/q^2\,,
\qquad \varepsilon_p = \pi_P/\Omega^2 \,.
\end{equation}
$\pi_{T,L,P}$ are defined by writing the photon self-energy
in the form
\begin{equation}\label{pidecomp}
\pi_{\mu\nu}(q) = \pi_T R_{\mu\nu} + \pi_L Q_{\mu\nu} + \pi_P P_{\mu\nu} \,.
\end{equation}
with
	\begin{eqnarray}
R_{\mu\nu} &=& g_{\mu\nu} - {q_\mu q_\nu \over q^2} - Q_{\mu\nu} \,,
\label{R} \\*
Q_{\mu\nu} &=& - {q^2 \over Q^2} \left( v_\mu - {\Omega q_\mu
\over q^2} \right) \left( v_\nu - {\Omega q_\nu
\over q^2} \right) \,, \label{Q} \\*
P_{\mu\nu} &=& {i \over Q} \epsilon_{\mu\nu\alpha\beta} q^\alpha
v^\beta \label{P} \,,
	\end{eqnarray}
where $v^\mu=(1, \vec 0)$ is the center-of-mass velocity of the medium.
Several useful properties of the tensors $R$, $Q$ and $P$ are
given explicitly in Ref.~\cite{pandcpodd}, to which we refer the
reader.  The solutions to Eqs.\ (\ref{phonshelltr}) and
(\ref{phonshelllong}), which we denote by $\Omega_s(Q)$, give
the energy-momentum relation for the three possible polarization
modes of the photon.  For non-chiral media $\pi_P$ arises only
through parity violation in weak interactions and hence must be
small.  Therefore, in what follows, we will neglect its effects,
thereby assuming that the two transverse degrees of freedom of
the photon are degenerate~\cite{footnote2}.

For future purposes, it is useful to recall that
the Eqs.\ (\ref{phonshelltr}) and (\ref{phonshelllong})
for the dispersion relations are equivalent to
\begin{equation}\label{photonshellpi}
\Omega_{t,l}^2 - Q^2 = \pi_{T,L} \,.
\end{equation}
In the literature, it is customary to use the indices of
refraction, which is yet another way of experssing
the dispersion relations.  Introducing the functions
\begin{equation}
\label{defngen}
n_{T,L}(\Omega,Q) = \sqrt{1 - \frac{\pi_{T,L}}{\Omega^2}}\,,
\end{equation}
solving Eq.~(\ref{photonshellpi}) is then equivalent to solve
\begin{equation}\label{photonshelln}
n_{T,L} = Q/\Omega_{t,l}\,.
\end{equation}
The indices of refraction are defined by
\begin{eqnarray}\label{defn}
n_{t,l} & \equiv & Q/\Omega_{t,l}(Q)\nonumber\\
& = & n_{T,L}(\Omega_{t,l},Q)\,,
\end{eqnarray}
so that the condition of Eq.\ (\ref{cerenkcond1}) is expressed as
$n_{t,l} > 1$.  Eq.~(\ref{phonshelltr}) implies the familiar relation
$n_{t} = \sqrt{\varepsilon_{t}(\Omega_{t}(Q),Q)}$.

\section{The $\nu\nu\gamma$ vertex}
In the following we follow closely the arguments and results
given in Ref.~\cite{indcharge}. We define the $\nu\nu\gamma$ amplitude by
\begin{equation}\label{amp}
M = -i\sqrt{N_s}\epsilon^{(s)\ast}_\mu (q)\overline u(k') \Gamma^\mu u(k) \,,
\end{equation}
where $\epsilon^{(s)}_\mu(q)$ is the polarization vector
of the emitted photon and the index $s$ indicates its polarization,
with $s = 1,2$ for the two (degenerate)
transverse modes and $s=3$ for the longitudinal one.
The factor $\sqrt{N_s}$ is necessary because
 the normalization of the photon wavefunction in the medium
is not the same as in the vacuum\cite{zaidi}.  Assuming that the polarization
vectors are normalized such that
\begin{equation}\label{polsum}
\sum_{s = 1,2} \epsilon^{(s)}_\mu\epsilon^{(s)}_\nu
= -R_{\mu\nu}
\end{equation}
and\footnote{In Ref.\ \cite{pandcpodd}, for example, the
right hand side of Eq.\ (\ref{polsuml}) has an extra minus sign, which
is the correct relation for timelike photons.  The present form
is appropriate for spacelike ones.}
\begin{equation}\label{polsuml}
\epsilon^{(3)}_\mu\epsilon^{(3)}_\nu = Q_{\mu\nu}\,,
\end{equation}
where $R_{\mu\nu}$ and $Q_{\mu\nu}$ have been defined in
Eqs.\ (\ref{R}) and (\ref{Q}), one obtains~\cite{pandcpodd,CNP88}
	\begin{eqnarray}
\label{normt}
N_{t,l} = \left.\left(
\frac{2\Omega}{\frac{\partial}{\partial\Omega}(\Omega^2 - \pi_{T,L} )}
\right)\right|_{\Omega = \Omega_{t,l}} \,,
	\end{eqnarray}
which, in terms of the index of refraction $n_{T,L}$, reads
	\begin{eqnarray}
N_{t,l} = \frac{n_{t,l}^{-1}}{n_{t,l} +
\left.\Omega_{t,l}\left(\frac{\partial n_{T,L}}{\partial\Omega}
\right)\right|_{{\Omega = \Omega_{t,l}}}}\,,
\end{eqnarray}

In Ref.~\cite{indcharge}, it has been shown that
the vertex function $\Gamma_\mu$ is given to leading order
in the Fermi constant by
\begin{equation}\label{vertex}
\Gamma_\mu = - \frac{\sqrt 2 G_F} {e}\gamma^\rho L
({\cal A}\pi_{\mu\rho} + {\cal B}\pi^5_{\mu\rho})\,,
\end{equation}
where $L={1\over2}(1-\gamma_5)$ is the projection operator for
left chirality, $\pi_{\mu\nu}$ is the photon self-energy and
$\pi^5_{\mu\nu}$ is a similarly defined function.  As shown in
Ref.~\cite{indcharge}, Eq.~(\ref{vertex}) is valid in all orders of the
electromagnetic interactions, where the most general form for
$\pi_{\mu\nu}$ is given in Eq.~(\ref{pidecomp}) while for
$\pi^5_{\mu\nu}$ it is
\begin{equation}\label{pi5decomp}
\pi^5_{\mu\rho} = \pi^5 P_{\mu\rho}\,.
\end{equation}
Finally, the constants ${\cal A}$ and
${\cal B}$ appearing in Eq.\ (\ref{vertex}) are defined by
writing the four-fermion interaction
between the neutrino and the electron in the form
\begin{equation}\label{Lweak}
{\cal L} {\rm _{int}^{(weak)}} =
-\sqrt 2 G_F \; [\bar \nu \gamma^\rho L\nu ] \;
[\bar f \gamma_\rho ({\cal A + B} \gamma_5) f ]\,,
\end{equation}
where $f$ stands for the electron field.  In the
standard model of electroweak interactions,
\begin{eqnarray}\label{A}
{\cal A} &=& \left\{ \begin{array}{ll}
2 \sin^2 \theta_W + {1 \over 2} \qquad & \mbox{for $\nu_e$} \\
2 \sin^2 \theta_W - {1 \over 2}  & \mbox{for $\nu_\mu, \nu_\tau$} \,.
\end{array} \right. \\*
{\cal B} &=& \left\{ \begin{array}{ll}
-\frac{1}{2}\qquad & \mbox{for $\nu_e$} \\
+\frac{1}{2}  & \mbox{for $\nu_\mu, \nu_\tau$} \,.
\end{array} \right. \label{B}
\end{eqnarray}

\section{Calculation of the rate}
The emission rate for transverse or longitudinal photons in a momentum range
from $Q_1$ to $Q_2$ is given by
	\begin{eqnarray}
\label{rateQ}
R_{t,l} &=& \frac{1}{16\pi \omega^2}\int_{Q_1}^{Q_2}
dQ \; \frac{Q}{\Omega_{t,l}(Q)}\left(\sum_s |M|^2\right) \\*
&=& \frac{1}{16\pi \omega}\int_{\xi_1}^{\xi_2}
d\xi \; n_{t,l} \left(\sum_s |M|^2\right) \,,
\label{ratexi}
	\end{eqnarray}
where in the last step, we have defined a new dimensionless
variable
	\begin{eqnarray}
\label{xi}
\xi\equiv \frac{Q}{\omega} \,.
	\end{eqnarray}
The limits of the integral in Eq.\ (\ref{ratexi}) are determined either
by our range of interest (e.g., we may be interested in just the
optical range), or by the range for which propagating photon modes exist.
If, however, propagating modes are available for the entire range of momenta
allowed by kinematics, the total rate will be given by
\begin{equation}\label{totratexi}
R_{t,l}^{\rm (tot)} = \frac{1}{16\pi \omega}\int_0^{\xi_{\rm max}}
d\xi \; n_{t,l} \left(\sum_s |M|^2\right)\,.
\end{equation}
Here, the upper limit of the integral should be
determined from the left inequality in Eq.\ (\ref{coslimits}),
and is given by
\begin{equation}\label{ximax}
\xi_{\rm max} = \frac{2n_{t,l}}{n_{t,l} + 1} \,.
\end{equation}
In general $n_{t,l}$ is a function of $Q$ and therefore of
$\xi$, so that Eq.~(\ref{ximax}) becomes
an implicit equation from which $\xi_{\rm max}$ is determined.


Whether we are interested in the total rate or the rate in any
particular range of momenta, the
result will be different for transverse and longitudinal photons
since they have different dispersion relations. Moreover, the
polarization sums in $\sum |M|^2$ will also involve different
polarization states. Thus, we carry out the calculation of the
rates for transverse and longitudinal photons separately.

\subsection{Transverse photons}
Using the relation in Eq.~(\ref{polsum}) for the polarization sum,
it then follows that
\begin{eqnarray}\label{M2}
\sum_{s = 1,2}|M|^2 & = & \frac{G_F^2}{4\pi\alpha} \; \omega^2
N_t \left(1 - \frac{1}{n_t^2}\right)
\left\{({\cal A}^2|\pi_T|^2 + {\cal B}^2|\pi^5|^2)\left[\left(2
- \frac{\xi}{n_t}\right)^2 + \xi^2\right]\right.\nonumber\\*
& & \qquad  - \left.4{\cal AB}\mbox{Re}\, \pi_T^\ast\pi^5\left(2\xi -
\frac{\xi^2}{n_t} \right)\right\}\,,
\end{eqnarray}
where we have neglected the contribution from $\pi_P$, as
already stated.  It must be remembered that in Eq.~(\ref{M2}) the
functions $\pi_T$ and $\pi^5$ are evaluated at the correct
photon dispersion relation, e.g., $\pi_T =
\pi_T(\Omega_t(Q),Q)$.  In order to carry out the integral
indicated in Eq.~(\ref{totratexi}) we need the explicit expressions for the
functions $\pi_T$ and $\pi^5$. We now argue as follows.  For an
electron gas, the 1-loop formulas for these two functions can be
inferred from the calculations of Ref.~\cite{dnp}.  In particular, it
follows that $\pi^5(0,0) = 0$, and motivated by this we
neglect the contribution from this function in Eq.~(\ref{M2}).  Then
we use
\begin{equation}\label{pisubt}
\pi_T = -\xi^2 \omega^2\left(1 - \frac{1}{n_t^2}\right)\,,
\end{equation}
which follows from Eqs.\ (\ref{photonshellpi}) and (\ref{defn}),
to obtain finally
\begin{equation}\label{ratefinal}
R_t = \frac{G_F^2 {\cal A}^2} {64\pi^2\alpha} \omega^5 F_t \,,
\end{equation}
where
\begin{equation}\label{Ft}
F_t \equiv  \int_{\xi_1}^{\xi_2}d\xi \; \xi^4 n_tN_t
 \left(1 - \frac{1}{n_t^2}\right)^3
\left[ \left( 2 - \frac{\xi}{n_t} \right)^2 + \xi^2 \right] \,.
\end{equation}
Eq.~(\ref{ratefinal}) is useful for making numerical estimates since
all the unknown aspects are contained in the single function
$F_t$. The evaluation of this function is involved because, as
mentioned above, the refractive index is a function of $Q$, and
therefore of $\xi$. However, to obtain a rough estimate of the
rates involved, we can pretend that $n_t$ is a constant over the
range of integration. In this
case we can perform the integral in Eq.\ (\ref{Ft}) for any
given value of $n_t$.

As an illustration, let us consider
the case of non-magnetic matter, i.e.,
materials for which the magnetic permeability $\mu = 1$,
or equivalently  $\varepsilon_t
= \varepsilon_l\equiv\varepsilon$. For large frequencies~\cite{LLbookEM},
\begin{equation}\label{eq64.4}
\varepsilon(\Omega)
= \varepsilon_\infty - \frac{\Omega_p^2}{\Omega^2}\,,
\end{equation}
where the asymptotic value $\varepsilon_\infty$ and
the plasma frequency $\Omega_p$ can be expressed
in terms of the imaginary part of the dielectric function
$\mbox{Im}\;\varepsilon$.  An important point to
notice is that the condition $\mbox{Im}\,\varepsilon > 0$
for $\Omega > 0$, which follows from fundamental
physical requirements \cite{stabcond},
implies that $\varepsilon_\infty > 1$ and,
therefore, at high frequencies, $n_t\approx\sqrt{\varepsilon_\infty} > 1$.
Thus, for example, if we assume that $n_t$ is constant at the value
$\sqrt{\varepsilon_\infty}$ within the range of integration for
which  $\xi_2=1$ and $\xi_1\ll\xi_2$,
 Eq.\ (\ref{Ft})
gives $F_t=(1 - n_t^{-2})^3 \times \left( {33 \over 35n_t} -
{2 \over 3n_t^2} + {1 \over 7n_t^3} \right)$.
For other values of $\xi_2$  the result can be read from Fig. 1.
As can be seen from that figure, $F_t$ becomes negligibly
small for any
value of $n_t$ if  $\xi_2$ is less than about 0.4.
The function increases rapidly as
the value of $\xi_2$ increases and, for values
of $\xi_2$ around 1 (which implies photon energies
of the order of the incident neutrino energy),
the function increases rapidly as the index of refraction
increases.
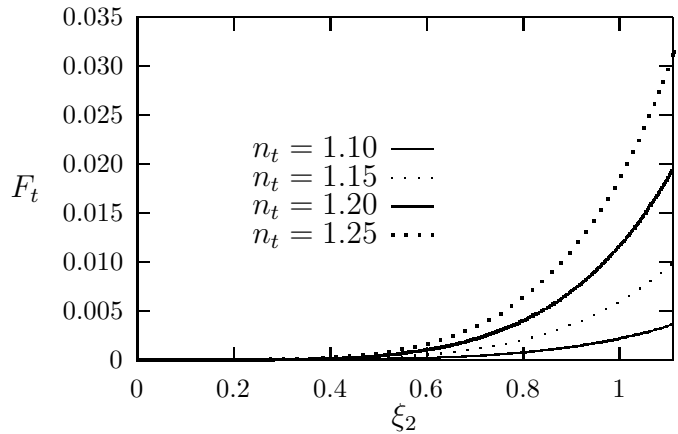
\begin{figure}
\begin{center}
\setlength{\unitlength}{0.240900pt}
\ifx\plotpoint\undefined\newsavebox{\plotpoint}\fi
\sbox{\plotpoint}{\rule[-0.200pt]{0.400pt}{0.400pt}}%
\begin{picture}(1125,675)(0,0)
\font\gnuplot=cmr10 at 10pt
\gnuplot
\sbox{\plotpoint}{\rule[-0.200pt]{0.400pt}{0.400pt}}%
\put(220.0,113.0){\rule[-0.200pt]{202.597pt}{0.400pt}}
\put(220.0,113.0){\rule[-0.200pt]{0.400pt}{129.845pt}}
\put(220.0,113.0){\rule[-0.200pt]{4.818pt}{0.400pt}}
\put(198,113){\makebox(0,0)[r]{0}}
\put(1041.0,113.0){\rule[-0.200pt]{4.818pt}{0.400pt}}
\put(220.0,190.0){\rule[-0.200pt]{4.818pt}{0.400pt}}
\put(198,190){\makebox(0,0)[r]{0.005}}
\put(1041.0,190.0){\rule[-0.200pt]{4.818pt}{0.400pt}}
\put(220.0,267.0){\rule[-0.200pt]{4.818pt}{0.400pt}}
\put(198,267){\makebox(0,0)[r]{0.010}}
\put(1041.0,267.0){\rule[-0.200pt]{4.818pt}{0.400pt}}
\put(220.0,344.0){\rule[-0.200pt]{4.818pt}{0.400pt}}
\put(198,344){\makebox(0,0)[r]{0.015}}
\put(1041.0,344.0){\rule[-0.200pt]{4.818pt}{0.400pt}}
\put(220.0,421.0){\rule[-0.200pt]{4.818pt}{0.400pt}}
\put(198,421){\makebox(0,0)[r]{0.020}}
\put(1041.0,421.0){\rule[-0.200pt]{4.818pt}{0.400pt}}
\put(220.0,498.0){\rule[-0.200pt]{4.818pt}{0.400pt}}
\put(198,498){\makebox(0,0)[r]{0.025}}
\put(1041.0,498.0){\rule[-0.200pt]{4.818pt}{0.400pt}}
\put(220.0,575.0){\rule[-0.200pt]{4.818pt}{0.400pt}}
\put(198,575){\makebox(0,0)[r]{0.030}}
\put(1041.0,575.0){\rule[-0.200pt]{4.818pt}{0.400pt}}
\put(220.0,652.0){\rule[-0.200pt]{4.818pt}{0.400pt}}
\put(198,652){\makebox(0,0)[r]{0.035}}
\put(1041.0,652.0){\rule[-0.200pt]{4.818pt}{0.400pt}}
\put(220.0,113.0){\rule[-0.200pt]{0.400pt}{4.818pt}}
\put(220,68){\makebox(0,0){0}}
\put(220.0,632.0){\rule[-0.200pt]{0.400pt}{4.818pt}}
\put(371.0,113.0){\rule[-0.200pt]{0.400pt}{4.818pt}}
\put(371,68){\makebox(0,0){0.2}}
\put(371.0,632.0){\rule[-0.200pt]{0.400pt}{4.818pt}}
\put(523.0,113.0){\rule[-0.200pt]{0.400pt}{4.818pt}}
\put(523,68){\makebox(0,0){0.4}}
\put(523.0,632.0){\rule[-0.200pt]{0.400pt}{4.818pt}}
\put(674.0,113.0){\rule[-0.200pt]{0.400pt}{4.818pt}}
\put(674,68){\makebox(0,0){0.6}}
\put(674.0,632.0){\rule[-0.200pt]{0.400pt}{4.818pt}}
\put(826.0,113.0){\rule[-0.200pt]{0.400pt}{4.818pt}}
\put(826,68){\makebox(0,0){0.8}}
\put(826.0,632.0){\rule[-0.200pt]{0.400pt}{4.818pt}}
\put(977.0,113.0){\rule[-0.200pt]{0.400pt}{4.818pt}}
\put(977,68){\makebox(0,0){1}}
\put(977.0,632.0){\rule[-0.200pt]{0.400pt}{4.818pt}}
\put(220.0,113.0){\rule[-0.200pt]{202.597pt}{0.400pt}}
\put(1061.0,113.0){\rule[-0.200pt]{0.400pt}{129.845pt}}
\put(220.0,652.0){\rule[-0.200pt]{202.597pt}{0.400pt}}
\put(45,382){\makebox(0,0){$F_t$}}
\put(640,23){\makebox(0,0){$\xi_2$}}
\put(220.0,113.0){\rule[-0.200pt]{0.400pt}{129.845pt}}
\put(598,444){\makebox(0,0)[r]{$n_t = 1.10$}}
\put(620.0,444.0){\rule[-0.200pt]{15.899pt}{0.400pt}}
\put(220,113){\usebox{\plotpoint}}
\put(526,112.67){\rule{1.927pt}{0.400pt}}
\multiput(526.00,112.17)(4.000,1.000){2}{\rule{0.964pt}{0.400pt}}
\put(220.0,113.0){\rule[-0.200pt]{73.715pt}{0.400pt}}
\put(611,113.67){\rule{1.927pt}{0.400pt}}
\multiput(611.00,113.17)(4.000,1.000){2}{\rule{0.964pt}{0.400pt}}
\put(534.0,114.0){\rule[-0.200pt]{18.549pt}{0.400pt}}
\put(653,114.67){\rule{2.168pt}{0.400pt}}
\multiput(653.00,114.17)(4.500,1.000){2}{\rule{1.084pt}{0.400pt}}
\put(619.0,115.0){\rule[-0.200pt]{8.191pt}{0.400pt}}
\put(687,115.67){\rule{2.168pt}{0.400pt}}
\multiput(687.00,115.17)(4.500,1.000){2}{\rule{1.084pt}{0.400pt}}
\put(662.0,116.0){\rule[-0.200pt]{6.022pt}{0.400pt}}
\put(713,116.67){\rule{1.927pt}{0.400pt}}
\multiput(713.00,116.17)(4.000,1.000){2}{\rule{0.964pt}{0.400pt}}
\put(696.0,117.0){\rule[-0.200pt]{4.095pt}{0.400pt}}
\put(730,117.67){\rule{1.927pt}{0.400pt}}
\multiput(730.00,117.17)(4.000,1.000){2}{\rule{0.964pt}{0.400pt}}
\put(721.0,118.0){\rule[-0.200pt]{2.168pt}{0.400pt}}
\put(755,118.67){\rule{2.168pt}{0.400pt}}
\multiput(755.00,118.17)(4.500,1.000){2}{\rule{1.084pt}{0.400pt}}
\put(764,119.67){\rule{1.927pt}{0.400pt}}
\multiput(764.00,119.17)(4.000,1.000){2}{\rule{0.964pt}{0.400pt}}
\put(738.0,119.0){\rule[-0.200pt]{4.095pt}{0.400pt}}
\put(781,120.67){\rule{1.927pt}{0.400pt}}
\multiput(781.00,120.17)(4.000,1.000){2}{\rule{0.964pt}{0.400pt}}
\put(772.0,121.0){\rule[-0.200pt]{2.168pt}{0.400pt}}
\put(798,121.67){\rule{1.927pt}{0.400pt}}
\multiput(798.00,121.17)(4.000,1.000){2}{\rule{0.964pt}{0.400pt}}
\put(806,122.67){\rule{2.168pt}{0.400pt}}
\multiput(806.00,122.17)(4.500,1.000){2}{\rule{1.084pt}{0.400pt}}
\put(789.0,122.0){\rule[-0.200pt]{2.168pt}{0.400pt}}
\put(823,123.67){\rule{2.168pt}{0.400pt}}
\multiput(823.00,123.17)(4.500,1.000){2}{\rule{1.084pt}{0.400pt}}
\put(832,124.67){\rule{1.927pt}{0.400pt}}
\multiput(832.00,124.17)(4.000,1.000){2}{\rule{0.964pt}{0.400pt}}
\put(840,125.67){\rule{2.168pt}{0.400pt}}
\multiput(840.00,125.17)(4.500,1.000){2}{\rule{1.084pt}{0.400pt}}
\put(849,126.67){\rule{1.927pt}{0.400pt}}
\multiput(849.00,126.17)(4.000,1.000){2}{\rule{0.964pt}{0.400pt}}
\put(857,127.67){\rule{2.168pt}{0.400pt}}
\multiput(857.00,127.17)(4.500,1.000){2}{\rule{1.084pt}{0.400pt}}
\put(866,128.67){\rule{1.927pt}{0.400pt}}
\multiput(866.00,128.17)(4.000,1.000){2}{\rule{0.964pt}{0.400pt}}
\put(874,129.67){\rule{2.168pt}{0.400pt}}
\multiput(874.00,129.17)(4.500,1.000){2}{\rule{1.084pt}{0.400pt}}
\put(883,130.67){\rule{1.927pt}{0.400pt}}
\multiput(883.00,130.17)(4.000,1.000){2}{\rule{0.964pt}{0.400pt}}
\put(891,131.67){\rule{2.168pt}{0.400pt}}
\multiput(891.00,131.17)(4.500,1.000){2}{\rule{1.084pt}{0.400pt}}
\put(900,132.67){\rule{1.927pt}{0.400pt}}
\multiput(900.00,132.17)(4.000,1.000){2}{\rule{0.964pt}{0.400pt}}
\put(908,133.67){\rule{2.168pt}{0.400pt}}
\multiput(908.00,133.17)(4.500,1.000){2}{\rule{1.084pt}{0.400pt}}
\put(917,135.17){\rule{1.700pt}{0.400pt}}
\multiput(917.00,134.17)(4.472,2.000){2}{\rule{0.850pt}{0.400pt}}
\put(925,136.67){\rule{2.168pt}{0.400pt}}
\multiput(925.00,136.17)(4.500,1.000){2}{\rule{1.084pt}{0.400pt}}
\put(934,138.17){\rule{1.700pt}{0.400pt}}
\multiput(934.00,137.17)(4.472,2.000){2}{\rule{0.850pt}{0.400pt}}
\put(942,139.67){\rule{2.168pt}{0.400pt}}
\multiput(942.00,139.17)(4.500,1.000){2}{\rule{1.084pt}{0.400pt}}
\put(951,141.17){\rule{1.700pt}{0.400pt}}
\multiput(951.00,140.17)(4.472,2.000){2}{\rule{0.850pt}{0.400pt}}
\put(959,142.67){\rule{2.168pt}{0.400pt}}
\multiput(959.00,142.17)(4.500,1.000){2}{\rule{1.084pt}{0.400pt}}
\put(968,144.17){\rule{1.700pt}{0.400pt}}
\multiput(968.00,143.17)(4.472,2.000){2}{\rule{0.850pt}{0.400pt}}
\put(976,146.17){\rule{1.900pt}{0.400pt}}
\multiput(976.00,145.17)(5.056,2.000){2}{\rule{0.950pt}{0.400pt}}
\put(985,148.17){\rule{1.700pt}{0.400pt}}
\multiput(985.00,147.17)(4.472,2.000){2}{\rule{0.850pt}{0.400pt}}
\put(993,150.17){\rule{1.900pt}{0.400pt}}
\multiput(993.00,149.17)(5.056,2.000){2}{\rule{0.950pt}{0.400pt}}
\put(1002,152.17){\rule{1.700pt}{0.400pt}}
\multiput(1002.00,151.17)(4.472,2.000){2}{\rule{0.850pt}{0.400pt}}
\put(1010,154.17){\rule{1.900pt}{0.400pt}}
\multiput(1010.00,153.17)(5.056,2.000){2}{\rule{0.950pt}{0.400pt}}
\put(1019,156.17){\rule{1.700pt}{0.400pt}}
\multiput(1019.00,155.17)(4.472,2.000){2}{\rule{0.850pt}{0.400pt}}
\multiput(1027.00,158.61)(1.802,0.447){3}{\rule{1.300pt}{0.108pt}}
\multiput(1027.00,157.17)(6.302,3.000){2}{\rule{0.650pt}{0.400pt}}
\put(1036,161.17){\rule{1.700pt}{0.400pt}}
\multiput(1036.00,160.17)(4.472,2.000){2}{\rule{0.850pt}{0.400pt}}
\multiput(1044.00,163.61)(1.802,0.447){3}{\rule{1.300pt}{0.108pt}}
\multiput(1044.00,162.17)(6.302,3.000){2}{\rule{0.650pt}{0.400pt}}
\multiput(1053.00,166.61)(1.579,0.447){3}{\rule{1.167pt}{0.108pt}}
\multiput(1053.00,165.17)(5.579,3.000){2}{\rule{0.583pt}{0.400pt}}
\put(815.0,124.0){\rule[-0.200pt]{1.927pt}{0.400pt}}
\put(598,399){\makebox(0,0)[r]{$n_t = 1.15$}}
\multiput(620,399)(20.756,0.000){4}{\usebox{\plotpoint}}
\put(220,113){\usebox{\plotpoint}}
\put(220.00,113.00){\usebox{\plotpoint}}
\multiput(228,113)(20.756,0.000){0}{\usebox{\plotpoint}}
\put(240.76,113.00){\usebox{\plotpoint}}
\multiput(245,113)(20.756,0.000){0}{\usebox{\plotpoint}}
\put(261.51,113.00){\usebox{\plotpoint}}
\multiput(262,113)(20.756,0.000){0}{\usebox{\plotpoint}}
\multiput(271,113)(20.756,0.000){0}{\usebox{\plotpoint}}
\put(282.27,113.00){\usebox{\plotpoint}}
\multiput(288,113)(20.756,0.000){0}{\usebox{\plotpoint}}
\put(303.02,113.00){\usebox{\plotpoint}}
\multiput(305,113)(20.756,0.000){0}{\usebox{\plotpoint}}
\multiput(313,113)(20.756,0.000){0}{\usebox{\plotpoint}}
\put(323.78,113.00){\usebox{\plotpoint}}
\multiput(330,113)(20.756,0.000){0}{\usebox{\plotpoint}}
\put(344.53,113.00){\usebox{\plotpoint}}
\multiput(347,113)(20.756,0.000){0}{\usebox{\plotpoint}}
\multiput(356,113)(20.756,0.000){0}{\usebox{\plotpoint}}
\put(365.29,113.00){\usebox{\plotpoint}}
\multiput(373,113)(20.756,0.000){0}{\usebox{\plotpoint}}
\put(386.04,113.00){\usebox{\plotpoint}}
\multiput(390,113)(20.756,0.000){0}{\usebox{\plotpoint}}
\put(406.80,113.00){\usebox{\plotpoint}}
\multiput(407,113)(20.756,0.000){0}{\usebox{\plotpoint}}
\multiput(415,113)(20.756,0.000){0}{\usebox{\plotpoint}}
\put(427.55,113.00){\usebox{\plotpoint}}
\multiput(432,113)(20.756,0.000){0}{\usebox{\plotpoint}}
\put(448.31,113.00){\usebox{\plotpoint}}
\multiput(449,113)(20.756,0.000){0}{\usebox{\plotpoint}}
\multiput(458,113)(20.756,0.000){0}{\usebox{\plotpoint}}
\put(469.05,113.34){\usebox{\plotpoint}}
\multiput(475,114)(20.756,0.000){0}{\usebox{\plotpoint}}
\put(489.77,114.00){\usebox{\plotpoint}}
\multiput(492,114)(20.756,0.000){0}{\usebox{\plotpoint}}
\multiput(500,114)(20.756,0.000){0}{\usebox{\plotpoint}}
\put(510.52,114.00){\usebox{\plotpoint}}
\multiput(517,114)(20.756,0.000){0}{\usebox{\plotpoint}}
\put(531.28,114.00){\usebox{\plotpoint}}
\multiput(534,114)(20.629,2.292){0}{\usebox{\plotpoint}}
\multiput(543,115)(20.756,0.000){0}{\usebox{\plotpoint}}
\put(551.98,115.00){\usebox{\plotpoint}}
\multiput(560,115)(20.756,0.000){0}{\usebox{\plotpoint}}
\put(572.70,115.52){\usebox{\plotpoint}}
\multiput(577,116)(20.756,0.000){0}{\usebox{\plotpoint}}
\put(593.43,116.00){\usebox{\plotpoint}}
\multiput(594,116)(20.595,2.574){0}{\usebox{\plotpoint}}
\multiput(602,117)(20.756,0.000){0}{\usebox{\plotpoint}}
\put(614.13,117.00){\usebox{\plotpoint}}
\multiput(619,117)(20.629,2.292){0}{\usebox{\plotpoint}}
\put(634.83,118.00){\usebox{\plotpoint}}
\multiput(636,118)(20.629,2.292){0}{\usebox{\plotpoint}}
\multiput(645,119)(20.595,2.574){0}{\usebox{\plotpoint}}
\put(655.46,120.00){\usebox{\plotpoint}}
\multiput(662,120)(20.595,2.574){0}{\usebox{\plotpoint}}
\put(676.12,121.68){\usebox{\plotpoint}}
\multiput(679,122)(20.756,0.000){0}{\usebox{\plotpoint}}
\multiput(687,122)(20.629,2.292){0}{\usebox{\plotpoint}}
\put(696.80,123.10){\usebox{\plotpoint}}
\multiput(704,124)(20.629,2.292){0}{\usebox{\plotpoint}}
\put(717.41,125.55){\usebox{\plotpoint}}
\multiput(721,126)(20.629,2.292){0}{\usebox{\plotpoint}}
\multiput(730,127)(20.595,2.574){0}{\usebox{\plotpoint}}
\put(738.02,128.00){\usebox{\plotpoint}}
\multiput(747,129)(20.136,5.034){0}{\usebox{\plotpoint}}
\put(758.45,131.38){\usebox{\plotpoint}}
\multiput(764,132)(20.595,2.574){0}{\usebox{\plotpoint}}
\put(778.94,134.54){\usebox{\plotpoint}}
\multiput(781,135)(20.136,5.034){0}{\usebox{\plotpoint}}
\multiput(789,137)(20.629,2.292){0}{\usebox{\plotpoint}}
\put(799.30,138.33){\usebox{\plotpoint}}
\multiput(806,140)(20.261,4.503){0}{\usebox{\plotpoint}}
\put(819.49,143.12){\usebox{\plotpoint}}
\multiput(823,144)(20.261,4.503){0}{\usebox{\plotpoint}}
\put(839.68,147.92){\usebox{\plotpoint}}
\multiput(840,148)(19.690,6.563){0}{\usebox{\plotpoint}}
\multiput(849,151)(20.136,5.034){0}{\usebox{\plotpoint}}
\put(859.56,153.85){\usebox{\plotpoint}}
\multiput(866,156)(19.434,7.288){0}{\usebox{\plotpoint}}
\put(879.29,160.18){\usebox{\plotpoint}}
\multiput(883,161)(19.434,7.288){0}{\usebox{\plotpoint}}
\put(898.69,167.42){\usebox{\plotpoint}}
\multiput(900,168)(19.434,7.288){0}{\usebox{\plotpoint}}
\multiput(908,171)(19.690,6.563){0}{\usebox{\plotpoint}}
\put(918.15,174.58){\usebox{\plotpoint}}
\multiput(925,178)(18.967,8.430){0}{\usebox{\plotpoint}}
\put(936.91,183.45){\usebox{\plotpoint}}
\multiput(942,186)(18.967,8.430){0}{\usebox{\plotpoint}}
\put(955.42,192.76){\usebox{\plotpoint}}
\multiput(959,195)(18.967,8.430){0}{\usebox{\plotpoint}}
\put(973.67,202.54){\usebox{\plotpoint}}
\multiput(976,204)(18.144,10.080){0}{\usebox{\plotpoint}}
\put(991.54,213.09){\usebox{\plotpoint}}
\multiput(993,214)(17.270,11.513){0}{\usebox{\plotpoint}}
\put(1008.57,224.93){\usebox{\plotpoint}}
\multiput(1010,226)(17.270,11.513){0}{\usebox{\plotpoint}}
\put(1025.53,236.89){\usebox{\plotpoint}}
\multiput(1027,238)(16.383,12.743){0}{\usebox{\plotpoint}}
\put(1042.01,249.51){\usebox{\plotpoint}}
\multiput(1044,251)(15.513,13.789){0}{\usebox{\plotpoint}}
\put(1057.68,263.10){\usebox{\plotpoint}}
\put(1061,266){\usebox{\plotpoint}}
\sbox{\plotpoint}{\rule[-0.400pt]{0.800pt}{0.800pt}}%
\put(598,354){\makebox(0,0)[r]{$n_t = 1.20$}}
\put(620.0,354.0){\rule[-0.400pt]{15.899pt}{0.800pt}}
\put(220,113){\usebox{\plotpoint}}
\put(432,111.84){\rule{2.168pt}{0.800pt}}
\multiput(432.00,111.34)(4.500,1.000){2}{\rule{1.084pt}{0.800pt}}
\put(220.0,113.0){\rule[-0.400pt]{51.071pt}{0.800pt}}
\put(492,112.84){\rule{1.927pt}{0.800pt}}
\multiput(492.00,112.34)(4.000,1.000){2}{\rule{0.964pt}{0.800pt}}
\put(441.0,114.0){\rule[-0.400pt]{12.286pt}{0.800pt}}
\put(526,113.84){\rule{1.927pt}{0.800pt}}
\multiput(526.00,113.34)(4.000,1.000){2}{\rule{0.964pt}{0.800pt}}
\put(500.0,115.0){\rule[-0.400pt]{6.263pt}{0.800pt}}
\put(543,114.84){\rule{1.927pt}{0.800pt}}
\multiput(543.00,114.34)(4.000,1.000){2}{\rule{0.964pt}{0.800pt}}
\put(534.0,116.0){\rule[-0.400pt]{2.168pt}{0.800pt}}
\put(568,115.84){\rule{2.168pt}{0.800pt}}
\multiput(568.00,115.34)(4.500,1.000){2}{\rule{1.084pt}{0.800pt}}
\put(577,116.84){\rule{1.927pt}{0.800pt}}
\multiput(577.00,116.34)(4.000,1.000){2}{\rule{0.964pt}{0.800pt}}
\put(551.0,117.0){\rule[-0.400pt]{4.095pt}{0.800pt}}
\put(594,117.84){\rule{1.927pt}{0.800pt}}
\multiput(594.00,117.34)(4.000,1.000){2}{\rule{0.964pt}{0.800pt}}
\put(602,118.84){\rule{2.168pt}{0.800pt}}
\multiput(602.00,118.34)(4.500,1.000){2}{\rule{1.084pt}{0.800pt}}
\put(611,119.84){\rule{1.927pt}{0.800pt}}
\multiput(611.00,119.34)(4.000,1.000){2}{\rule{0.964pt}{0.800pt}}
\put(585.0,119.0){\rule[-0.400pt]{2.168pt}{0.800pt}}
\put(628,120.84){\rule{1.927pt}{0.800pt}}
\multiput(628.00,120.34)(4.000,1.000){2}{\rule{0.964pt}{0.800pt}}
\put(636,122.34){\rule{2.168pt}{0.800pt}}
\multiput(636.00,121.34)(4.500,2.000){2}{\rule{1.084pt}{0.800pt}}
\put(645,123.84){\rule{1.927pt}{0.800pt}}
\multiput(645.00,123.34)(4.000,1.000){2}{\rule{0.964pt}{0.800pt}}
\put(653,124.84){\rule{2.168pt}{0.800pt}}
\multiput(653.00,124.34)(4.500,1.000){2}{\rule{1.084pt}{0.800pt}}
\put(662,125.84){\rule{1.927pt}{0.800pt}}
\multiput(662.00,125.34)(4.000,1.000){2}{\rule{0.964pt}{0.800pt}}
\put(670,127.34){\rule{2.168pt}{0.800pt}}
\multiput(670.00,126.34)(4.500,2.000){2}{\rule{1.084pt}{0.800pt}}
\put(679,128.84){\rule{1.927pt}{0.800pt}}
\multiput(679.00,128.34)(4.000,1.000){2}{\rule{0.964pt}{0.800pt}}
\put(687,130.34){\rule{2.168pt}{0.800pt}}
\multiput(687.00,129.34)(4.500,2.000){2}{\rule{1.084pt}{0.800pt}}
\put(696,131.84){\rule{1.927pt}{0.800pt}}
\multiput(696.00,131.34)(4.000,1.000){2}{\rule{0.964pt}{0.800pt}}
\put(704,133.34){\rule{2.168pt}{0.800pt}}
\multiput(704.00,132.34)(4.500,2.000){2}{\rule{1.084pt}{0.800pt}}
\put(713,135.34){\rule{1.927pt}{0.800pt}}
\multiput(713.00,134.34)(4.000,2.000){2}{\rule{0.964pt}{0.800pt}}
\put(721,137.34){\rule{2.168pt}{0.800pt}}
\multiput(721.00,136.34)(4.500,2.000){2}{\rule{1.084pt}{0.800pt}}
\put(730,139.34){\rule{1.927pt}{0.800pt}}
\multiput(730.00,138.34)(4.000,2.000){2}{\rule{0.964pt}{0.800pt}}
\put(738,141.84){\rule{2.168pt}{0.800pt}}
\multiput(738.00,140.34)(4.500,3.000){2}{\rule{1.084pt}{0.800pt}}
\put(747,144.34){\rule{1.927pt}{0.800pt}}
\multiput(747.00,143.34)(4.000,2.000){2}{\rule{0.964pt}{0.800pt}}
\put(755,146.84){\rule{2.168pt}{0.800pt}}
\multiput(755.00,145.34)(4.500,3.000){2}{\rule{1.084pt}{0.800pt}}
\put(764,149.84){\rule{1.927pt}{0.800pt}}
\multiput(764.00,148.34)(4.000,3.000){2}{\rule{0.964pt}{0.800pt}}
\put(772,152.84){\rule{2.168pt}{0.800pt}}
\multiput(772.00,151.34)(4.500,3.000){2}{\rule{1.084pt}{0.800pt}}
\put(781,155.84){\rule{1.927pt}{0.800pt}}
\multiput(781.00,154.34)(4.000,3.000){2}{\rule{0.964pt}{0.800pt}}
\put(789,158.84){\rule{2.168pt}{0.800pt}}
\multiput(789.00,157.34)(4.500,3.000){2}{\rule{1.084pt}{0.800pt}}
\put(798,162.34){\rule{1.800pt}{0.800pt}}
\multiput(798.00,160.34)(4.264,4.000){2}{\rule{0.900pt}{0.800pt}}
\put(806,165.84){\rule{2.168pt}{0.800pt}}
\multiput(806.00,164.34)(4.500,3.000){2}{\rule{1.084pt}{0.800pt}}
\put(815,169.34){\rule{1.800pt}{0.800pt}}
\multiput(815.00,167.34)(4.264,4.000){2}{\rule{0.900pt}{0.800pt}}
\put(823,173.34){\rule{2.000pt}{0.800pt}}
\multiput(823.00,171.34)(4.849,4.000){2}{\rule{1.000pt}{0.800pt}}
\multiput(832.00,178.38)(0.928,0.560){3}{\rule{1.480pt}{0.135pt}}
\multiput(832.00,175.34)(4.928,5.000){2}{\rule{0.740pt}{0.800pt}}
\put(840,182.34){\rule{2.000pt}{0.800pt}}
\multiput(840.00,180.34)(4.849,4.000){2}{\rule{1.000pt}{0.800pt}}
\multiput(849.00,187.38)(0.928,0.560){3}{\rule{1.480pt}{0.135pt}}
\multiput(849.00,184.34)(4.928,5.000){2}{\rule{0.740pt}{0.800pt}}
\multiput(857.00,192.38)(1.096,0.560){3}{\rule{1.640pt}{0.135pt}}
\multiput(857.00,189.34)(5.596,5.000){2}{\rule{0.820pt}{0.800pt}}
\multiput(866.00,197.38)(0.928,0.560){3}{\rule{1.480pt}{0.135pt}}
\multiput(866.00,194.34)(4.928,5.000){2}{\rule{0.740pt}{0.800pt}}
\multiput(874.00,202.39)(0.797,0.536){5}{\rule{1.400pt}{0.129pt}}
\multiput(874.00,199.34)(6.094,6.000){2}{\rule{0.700pt}{0.800pt}}
\multiput(883.00,208.39)(0.685,0.536){5}{\rule{1.267pt}{0.129pt}}
\multiput(883.00,205.34)(5.371,6.000){2}{\rule{0.633pt}{0.800pt}}
\multiput(891.00,214.39)(0.797,0.536){5}{\rule{1.400pt}{0.129pt}}
\multiput(891.00,211.34)(6.094,6.000){2}{\rule{0.700pt}{0.800pt}}
\multiput(900.00,220.40)(0.562,0.526){7}{\rule{1.114pt}{0.127pt}}
\multiput(900.00,217.34)(5.687,7.000){2}{\rule{0.557pt}{0.800pt}}
\multiput(908.00,227.39)(0.797,0.536){5}{\rule{1.400pt}{0.129pt}}
\multiput(908.00,224.34)(6.094,6.000){2}{\rule{0.700pt}{0.800pt}}
\multiput(917.00,233.40)(0.562,0.526){7}{\rule{1.114pt}{0.127pt}}
\multiput(917.00,230.34)(5.687,7.000){2}{\rule{0.557pt}{0.800pt}}
\multiput(925.00,240.40)(0.554,0.520){9}{\rule{1.100pt}{0.125pt}}
\multiput(925.00,237.34)(6.717,8.000){2}{\rule{0.550pt}{0.800pt}}
\multiput(934.00,248.40)(0.481,0.520){9}{\rule{1.000pt}{0.125pt}}
\multiput(934.00,245.34)(5.924,8.000){2}{\rule{0.500pt}{0.800pt}}
\multiput(942.00,256.40)(0.554,0.520){9}{\rule{1.100pt}{0.125pt}}
\multiput(942.00,253.34)(6.717,8.000){2}{\rule{0.550pt}{0.800pt}}
\multiput(952.40,263.00)(0.520,0.554){9}{\rule{0.125pt}{1.100pt}}
\multiput(949.34,263.00)(8.000,6.717){2}{\rule{0.800pt}{0.550pt}}
\multiput(959.00,273.40)(0.485,0.516){11}{\rule{1.000pt}{0.124pt}}
\multiput(959.00,270.34)(6.924,9.000){2}{\rule{0.500pt}{0.800pt}}
\multiput(969.40,281.00)(0.520,0.554){9}{\rule{0.125pt}{1.100pt}}
\multiput(966.34,281.00)(8.000,6.717){2}{\rule{0.800pt}{0.550pt}}
\multiput(977.40,290.00)(0.516,0.548){11}{\rule{0.124pt}{1.089pt}}
\multiput(974.34,290.00)(9.000,7.740){2}{\rule{0.800pt}{0.544pt}}
\multiput(986.40,300.00)(0.520,0.627){9}{\rule{0.125pt}{1.200pt}}
\multiput(983.34,300.00)(8.000,7.509){2}{\rule{0.800pt}{0.600pt}}
\multiput(994.40,310.00)(0.516,0.611){11}{\rule{0.124pt}{1.178pt}}
\multiput(991.34,310.00)(9.000,8.555){2}{\rule{0.800pt}{0.589pt}}
\multiput(1003.40,321.00)(0.520,0.700){9}{\rule{0.125pt}{1.300pt}}
\multiput(1000.34,321.00)(8.000,8.302){2}{\rule{0.800pt}{0.650pt}}
\multiput(1011.40,332.00)(0.516,0.674){11}{\rule{0.124pt}{1.267pt}}
\multiput(1008.34,332.00)(9.000,9.371){2}{\rule{0.800pt}{0.633pt}}
\multiput(1020.40,344.00)(0.520,0.847){9}{\rule{0.125pt}{1.500pt}}
\multiput(1017.34,344.00)(8.000,9.887){2}{\rule{0.800pt}{0.750pt}}
\multiput(1028.40,357.00)(0.516,0.674){11}{\rule{0.124pt}{1.267pt}}
\multiput(1025.34,357.00)(9.000,9.371){2}{\rule{0.800pt}{0.633pt}}
\multiput(1037.40,369.00)(0.520,0.920){9}{\rule{0.125pt}{1.600pt}}
\multiput(1034.34,369.00)(8.000,10.679){2}{\rule{0.800pt}{0.800pt}}
\multiput(1045.40,383.00)(0.516,0.800){11}{\rule{0.124pt}{1.444pt}}
\multiput(1042.34,383.00)(9.000,11.002){2}{\rule{0.800pt}{0.722pt}}
\multiput(1054.40,397.00)(0.520,0.993){9}{\rule{0.125pt}{1.700pt}}
\multiput(1051.34,397.00)(8.000,11.472){2}{\rule{0.800pt}{0.850pt}}
\put(619.0,122.0){\rule[-0.400pt]{2.168pt}{0.800pt}}
\sbox{\plotpoint}{\rule[-0.500pt]{1.000pt}{1.000pt}}%
\put(598,309){\makebox(0,0)[r]{$n_t = 1.25$}}
\multiput(620,309)(20.756,0.000){4}{\usebox{\plotpoint}}
\put(220,113){\usebox{\plotpoint}}
\put(220.00,113.00){\usebox{\plotpoint}}
\multiput(228,113)(20.756,0.000){0}{\usebox{\plotpoint}}
\put(240.76,113.00){\usebox{\plotpoint}}
\multiput(245,113)(20.756,0.000){0}{\usebox{\plotpoint}}
\put(261.51,113.00){\usebox{\plotpoint}}
\multiput(262,113)(20.756,0.000){0}{\usebox{\plotpoint}}
\multiput(271,113)(20.756,0.000){0}{\usebox{\plotpoint}}
\put(282.27,113.00){\usebox{\plotpoint}}
\multiput(288,113)(20.756,0.000){0}{\usebox{\plotpoint}}
\put(303.02,113.00){\usebox{\plotpoint}}
\multiput(305,113)(20.756,0.000){0}{\usebox{\plotpoint}}
\multiput(313,113)(20.756,0.000){0}{\usebox{\plotpoint}}
\put(323.78,113.00){\usebox{\plotpoint}}
\multiput(330,113)(20.756,0.000){0}{\usebox{\plotpoint}}
\put(344.53,113.00){\usebox{\plotpoint}}
\multiput(347,113)(20.756,0.000){0}{\usebox{\plotpoint}}
\multiput(356,113)(20.756,0.000){0}{\usebox{\plotpoint}}
\put(365.29,113.00){\usebox{\plotpoint}}
\multiput(373,113)(20.756,0.000){0}{\usebox{\plotpoint}}
\put(386.04,113.00){\usebox{\plotpoint}}
\multiput(390,113)(20.756,0.000){0}{\usebox{\plotpoint}}
\put(406.80,113.00){\usebox{\plotpoint}}
\multiput(407,113)(20.756,0.000){0}{\usebox{\plotpoint}}
\multiput(415,113)(20.629,2.292){0}{\usebox{\plotpoint}}
\put(427.50,114.00){\usebox{\plotpoint}}
\multiput(432,114)(20.756,0.000){0}{\usebox{\plotpoint}}
\put(448.26,114.00){\usebox{\plotpoint}}
\multiput(449,114)(20.756,0.000){0}{\usebox{\plotpoint}}
\multiput(458,114)(20.756,0.000){0}{\usebox{\plotpoint}}
\put(468.99,114.33){\usebox{\plotpoint}}
\multiput(475,115)(20.756,0.000){0}{\usebox{\plotpoint}}
\put(489.71,115.00){\usebox{\plotpoint}}
\multiput(492,115)(20.595,2.574){0}{\usebox{\plotpoint}}
\multiput(500,116)(20.756,0.000){0}{\usebox{\plotpoint}}
\put(510.40,116.00){\usebox{\plotpoint}}
\multiput(517,116)(20.629,2.292){0}{\usebox{\plotpoint}}
\put(531.10,117.00){\usebox{\plotpoint}}
\multiput(534,117)(20.629,2.292){0}{\usebox{\plotpoint}}
\multiput(543,118)(20.595,2.574){0}{\usebox{\plotpoint}}
\put(551.74,119.00){\usebox{\plotpoint}}
\multiput(560,119)(20.595,2.574){0}{\usebox{\plotpoint}}
\put(572.41,120.49){\usebox{\plotpoint}}
\multiput(577,121)(20.595,2.574){0}{\usebox{\plotpoint}}
\put(593.02,122.89){\usebox{\plotpoint}}
\multiput(594,123)(20.595,2.574){0}{\usebox{\plotpoint}}
\multiput(602,124)(20.629,2.292){0}{\usebox{\plotpoint}}
\put(613.58,125.64){\usebox{\plotpoint}}
\multiput(619,127)(20.629,2.292){0}{\usebox{\plotpoint}}
\put(633.93,129.48){\usebox{\plotpoint}}
\multiput(636,130)(20.629,2.292){0}{\usebox{\plotpoint}}
\multiput(645,131)(20.136,5.034){0}{\usebox{\plotpoint}}
\put(654.29,133.29){\usebox{\plotpoint}}
\multiput(662,135)(20.136,5.034){0}{\usebox{\plotpoint}}
\put(674.50,138.00){\usebox{\plotpoint}}
\multiput(679,139)(19.434,7.288){0}{\usebox{\plotpoint}}
\put(694.42,143.65){\usebox{\plotpoint}}
\multiput(696,144)(19.434,7.288){0}{\usebox{\plotpoint}}
\multiput(704,147)(19.690,6.563){0}{\usebox{\plotpoint}}
\put(714.03,150.39){\usebox{\plotpoint}}
\multiput(721,153)(18.967,8.430){0}{\usebox{\plotpoint}}
\put(733.25,158.22){\usebox{\plotpoint}}
\multiput(738,160)(18.967,8.430){0}{\usebox{\plotpoint}}
\put(752.21,166.61){\usebox{\plotpoint}}
\multiput(755,168)(18.967,8.430){0}{\usebox{\plotpoint}}
\put(770.61,176.13){\usebox{\plotpoint}}
\multiput(772,177)(18.967,8.430){0}{\usebox{\plotpoint}}
\put(788.86,185.91){\usebox{\plotpoint}}
\multiput(789,186)(17.270,11.513){0}{\usebox{\plotpoint}}
\multiput(798,192)(17.601,11.000){0}{\usebox{\plotpoint}}
\put(806.28,197.19){\usebox{\plotpoint}}
\put(822.73,209.77){\usebox{\plotpoint}}
\multiput(823,210)(17.270,11.513){0}{\usebox{\plotpoint}}
\put(839.21,222.31){\usebox{\plotpoint}}
\multiput(840,223)(15.513,13.789){0}{\usebox{\plotpoint}}
\put(854.42,236.42){\usebox{\plotpoint}}
\multiput(857,239)(15.513,13.789){0}{\usebox{\plotpoint}}
\put(869.58,250.58){\usebox{\plotpoint}}
\multiput(874,255)(13.885,15.427){0}{\usebox{\plotpoint}}
\put(883.70,265.79){\usebox{\plotpoint}}
\put(897.54,281.26){\usebox{\plotpoint}}
\multiput(900,284)(12.208,16.786){0}{\usebox{\plotpoint}}
\put(910.20,297.69){\usebox{\plotpoint}}
\put(922.89,314.10){\usebox{\plotpoint}}
\multiput(925,317)(12.453,16.604){0}{\usebox{\plotpoint}}
\put(935.14,330.84){\usebox{\plotpoint}}
\put(946.36,348.30){\usebox{\plotpoint}}
\put(957.25,365.94){\usebox{\plotpoint}}
\put(967.87,383.78){\usebox{\plotpoint}}
\multiput(968,384)(9.767,18.314){0}{\usebox{\plotpoint}}
\put(977.71,402.05){\usebox{\plotpoint}}
\put(987.51,420.33){\usebox{\plotpoint}}
\put(996.52,439.03){\usebox{\plotpoint}}
\put(1005.45,457.76){\usebox{\plotpoint}}
\put(1014.09,476.63){\usebox{\plotpoint}}
\put(1022.45,495.62){\usebox{\plotpoint}}
\put(1030.35,514.81){\usebox{\plotpoint}}
\put(1038.19,534.02){\usebox{\plotpoint}}
\multiput(1044,550)(7.563,19.328){2}{\usebox{\plotpoint}}
\put(1059.76,592.42){\usebox{\plotpoint}}
\put(1061,596){\usebox{\plotpoint}}
\end{picture}
\caption{\sf Plot of $F_t$
vs. $\xi_2$ for various values of $n_t$, assuming a constant
index of refraction and $\xi_1=0$.
Although we have plotted the function
over a common range of values $0 \leq \xi_2 \leq 1.11$,
it should be noted that, for a given value of $n_t$, the allowed
range of values of $\xi_2$ is limited by  Eq.\ (\protect\ref{ximax}).}
\end{center}
\end{figure}

\subsection{Longitudinal photons}
For longitudinal photons, the formulas are
analogous to those for transverse ones, with some obvious
substitutions like replacing $n_t$ by $n_l$. There is
no polarization sum now, and
we use Eq.~(\ref{polsuml}).
Then, using instead of Eq.\ (\ref{pisubt})
the relation appropriate
for the longitudinal photons,
\begin{equation}\label{pisubl}
\pi_L = -\xi^2 \omega^2\left(1 - \frac{1}{n_l^2}\right)\,,
\end{equation}
we obtain a formula for the rate which is
similar to that in Eq.\ (\ref{ratefinal}),
but with $F_t$ replaced by
\begin{equation}\label{Fl}
F_l \equiv \int_{\xi_1}^{\xi_2}d\xi \; \xi^4 n_lN_l
 \left(1 - \frac{1}{n_l^2}\right)^3
\left[ \left(2 - \frac{\xi}{n_l}\right)^2 - \xi^2\right] \,.
\end{equation}

While the expressions for the longitudinal and
transverse photons look similar, the
longitudinal photons behave very different from the transverse ones.
The dependence of the frequency on the wave vector
is given by Eq.\ (\ref{phonshelllong}).  It turns
out that for values of the momentum of the
order, or larger than, the inverse
Debye screening length $\lambda_D$, the
real and imaginary parts of $\Omega_l$ are
comparable\cite{LLbookPK}. Thus, above those photon
momenta, the longitudinal photon modes
do not exist.  Since $\lambda_D$ is of the
order of the Bohr radius, then for neutrino
energies of the order of an MeV we actually
have, for longitudinal photons,
$\xi_{{\rm max}} \sim 10^{-4}$.  On
the scale of Fig. 1, this gives a negligible
value of $F_l$, of order $\xi_{{\rm max}}^5$.

\section{Numerical estimates}
We now estimate the number of Cherenkov photons that will result
from the formulas above. For a flux $I$ of neutrinos, the number
of Cherenkov events occurring in a time $T$ in a detector of
volume $V$ is given by
	\begin{eqnarray}
{\cal N} = \frac{VTIR}{v} \,,
	\end{eqnarray}
where $R$ is the rate calculated in the last section and $v=c$
is the velocity of the neutrino.   This
gives
	\begin{eqnarray}
\label{numrate}
{\cal N}_{t,l} &=& 7.6 \times 10^9 {\cal A}^2 F_{t,l} \times \left(
{\omega \over 1\, {\rm MeV}} \right)^5
\left( {V\over 1 \, {\rm m}^3} \right)
\left( {T\over 1 \, {\rm day}} \right)
\left( {I \over I_\odot} \right) \,,
\label{N}
	\end{eqnarray}
where $I_\odot$ is the solar neutrino flux, $6\times
10^{10}\,{\rm cm^{-2}s^{-1}}$.
While it may seem from Eq.\ (\ref{N}) that
the number of Cherenkov events increase simply as
$\omega^5$, this is not really true since $F_{t,l}$ are also
functions of $\omega$ implicitly through $\xi$.

For optical photons, the formula above predicts a very small rate
since in this range, $Q_2\approx 3\,\mbox{eV}$. Thus, for example,
for a neutrino energy $\omega\approx 1\,\mbox{MeV}$, we have
$\xi_2\approx 3\times 10^{-6}$ so that
$\xi$ remains very small over the range of integration.
Assuming that the index of refraction is of order one,
$F_t\sim \xi_2^5\sim 10^{-28}$.  In general, if the
upper range of $Q$ is much smaller than $\omega$, then
$F_{t,l} \sim (Q_2/\omega)^5$, so that the rate
in Eq.~(\ref{numrate}) scales as $(Q_2/1\,\mbox{MeV})^5$.

Thus, for optical photons,
the effects considered by Grimus
and Neufeld \cite{grimus} seem to be much larger, at least if the
neutrino has a magnetic moment anywhere near the present experimental
limit. However, this should not be discouraging
because there are two important points to be remembered.

First, the effect considered
in Ref.~\cite{grimus} hinges on the assumption that the neutrinos
have an intrinsic magnetic moment of order $10^{-10}\mu_B$.
On the contrary, the effect we have discussed in the present paper
does not depend on any assumption about the
neutrino properties and/or interactions beyond those
specified by the Standard Model.

Secondly, the rate calculated by us
for transverse photons has a very different dependence on
neutrino energy than the rate calculated by Grimus and Neufeld
\cite{grimus}. In fact, using Eq.\ (18) of their paper, we can
easily deduce the ratio of the two effects assuming, for
illustration, that the
refractive index in roughly constant and that $\xi_2\gg\xi_1$:
	\begin{eqnarray}
{{\cal N}_{t} \over {\cal N}_{\rm mag}} = 1.2 \left(\frac{n_{t}}
{n_{t}^2 -1}\right)^2
{\cal A}^2 F_{t}
\left( {\omega \over 1\, {\rm MeV}} \right)^2 \left( {\mu \over
10^{-10} \mu_B} \right)^{-2} \,.
	\end{eqnarray}
If we do not restrict ourselves to look only
for photons in the optical range, but consider
instead photons with energies that span all the
kinematically allowed range, then
it is easily seen that even at neutrino energies around 1~MeV,
the effect we described in this paper is as important as the magnetic
moment effects even if the magnetic moment is close to its present
upper limit. If the magnetic moment is much lower, as
indicated by considerations on the neutrino flux from the supernova
SN1987A \cite{SN1987}, then of course the effect we described is much stronger.
In any case, as we emphasized before, the present effect is
predicted from the physics of the Standard Model only, and therefore
must be there.  In this sense, the
results of our calculations represent a firm prediction
if an experiment can be set up.  In fact, any
deviation from this prediction can be taken as
a serious indication that some
of the neutrino properties and/or interactions
are not the ones given by the Standard Model.

\paragraph*{Acknowledgements~:~}
The work of JCD was partially supported by Grant No. DGAPA-IN100691, and
that of JFN by  the US National
Science Foundation Grant PHY-9320692. The work of PBP was supported by
the Department of Science and Technology of the Government of India.


\begin{thebibliography}{99}
\bibitem{footnote1} Even if the neutrinos are massless in vacuum,
their dispersion relation in the medium is not the vacuum one
$\omega = K$ implied by Eq.~(\ref{nuonshell}).  While this is a well known
effect by now, for our present purposes it can be ignored.
\end{thebibliography}
\end{document}